\documentclass[10pt,journal,compsoc]{IEEEtran}

\ifCLASSOPTIONcompsoc
  \usepackage[nocompress]{cite}
\else
  \usepackage{cite}
\fi

\ifCLASSINFOpdf
   \usepackage[pdftex]{graphicx}
\else
\fi

\usepackage{underscore}
\usepackage[document]{ragged2e}
\usepackage[ruled,vlined]{algorithm2e}
\usepackage{dblfloatfix}
\usepackage{caption}
\usepackage{hyperref}
\usepackage[all]{hypcap}
\usepackage[T1]{fontenc}
\DeclareUnicodeCharacter{0301}{\'{e}}

\begin{document}
\title{An Overview of Forks and Coordination in Blockchain Development}
\author{Neo~C.K.~Yiu,~\IEEEmembership{Member,~IEEE\\ Department~of~Computer~Science,~University~of~Oxford\\neo-chungkit.yiu@kellogg.ox.ac.uk}
\IEEEcompsocitemizethanks{\IEEEcompsocthanksitem Neo C.K. Yiu was with Global Blockchain Advisory Board and Cyber Research Group, EC-Council}
\thanks{Manuscript first submitted for preprint on Feb 18, 2021.}}

\IEEEtitleabstractindextext{
\justify
\begin{abstract}
Blockchain is a continuously developing technology that has made digital transactions and related computing operations more transparent and secure through globally distributed and decentralized management of states, as well as the strong immutability of blocks mined and transactions validated in a network enabled by the blockchain technology. This manuscript is aimed at elaborating the concept of blockchain technology alongside its coordination and implementation with other emerging technologies, such as smart contract, which works with different blockchain frameworks, as well as enabling anonymous transactions and decentralized consensus amongst different untrusting parties. The discussion of blockchain forks is also covered in this manuscript, depicting fork events created in the blockchain process, their brief history, types, and impacts upon the blockchain development and operation.
\end{abstract}

}

\maketitle
\IEEEdisplaynontitleabstractindextext
\justifying
\IEEEpeerreviewmaketitle

\ifCLASSOPTIONcompsoc
\IEEEraisesectionheading{\section{Introduction}\label{sec:introduction}}
\else
\section{Introduction}
\label{sec:introduction}
\fi

\IEEEPARstart{T}{h}e blockchain is best described as decentralized value-exchange protocols or distributed ledgers that aim at decentralized data validation transparency via decentralized consensus reached in a network of distributed entities. These distributed entities would run their own node in a blockchain network individually, to validate transactions and even mine blocks in the network. As such, these distributed entities are termed as block miners or crypto miners of a chosen blockchain network with a specific consensus algorithm; for instance, Bitcoin miners compete to solve the "\emph{puzzles}" by being the first to calculate the block header hash, and to mine a series of hexadecimal valued blocks based on a consensus algorithm, namely the \emph{Proof-of-Work} (PoW) as elaborated in \cite{gervais2016security}. The consensus protocol and design of the network require a decentralized consensus to be reached amongst the participating nodes, so as to transit any state of a block, ledger or further to the entire network, with new transactions included in any newly mined block. The blockchain network and its primary implementation in the authentication and management of cryptocurrency is either permissionless and publicly distributed across the globe, or permissioned and incorporated into an enterprise setting with a concept of enterprise consortium.

Due to the nature of immutability in blockchain technology, the state of ledgers cannot be altered retroactively as every mined block does contain a hash pointer of its preceding block, in its block header, implying that any change could affect states of the entire blockchain network and will be easily identifiable. The immutability allows block miners to verify the ledger independently, easily, and autonomously in the peer-to-peer network. The distributed network is managed using a distributed timestamping server, which removes the possibility of multiple reproductions of unit values; for example, each unit of value could be transferred only once as explained in \cite{1}. Some of the significant characteristics \cite{2} in blockchain and cryptocurrency processing are listed below, with Bitcoin as the operational example.

\begin{itemize}
  \item \emph{Transaction}: Cryptocurrencies, such as Bitcoin, is a chain of digital signatures that could be transferred from one owner to the other in the network digitally with the help of a hash of its previous transaction, along with the public key of the next owner incorporated in it. These values are generally verified for certifying ownership of the crypto assets. To solve the issue of double spending, where a certain \emph{UTXO} (Unspent Transaction Outputs) of Bitcoin is spent more than once to pay different entities simultaneously, a single history of transaction receipts is publicly announced, and at the time of each transaction processing, the majority of nodes agreed a fact that the history of transactions is the one broadcast across the chain originally. The related Bitcoin miners and community are roped in as the trusted sources to validate the authenticity by verifying the initial history of the transactions along the canonical chain \cite{1}.
  \item \emph{Timestamp Server}: It works as a verifier that takes a hash and timestamps, and publishes them, thus proving a fact that the transaction data must have existed and a specific block must have been mined at the time. Each timestamp is included in its corresponding block header, and the hash of such block header is included in the block header of the next mined block, thus forming a "\emph{chain}".
  \item \emph{Decentralized Consensus}: This is an important and widely known type of consensus mechanism that involves calculating for a value that, when hashed, implements a distributed timestamp server on a peer-to-peer basis. By increasing a nonce value, based on the difficulty set, in the block until a value is found that gives the block header hash the required zero bits, the consensus algorithms, such as \cite{Proof-of-Work} could therefore be satisfied. Another important type of consensus mechanism is the \emph{Proof-of-Stake} \cite{bentov2014proof}, currently applied in \emph{Eth 2.0}, that helps achieve distributed consensus by choosing the creator node of the next block in the chain via various combinations of random data related to the stakes one holds.
  \item \emph{Merkle Tree}: Once a transaction is completed in a chained process, it could end up involving many previous blocks. To reclaim the disk space, the previous or spent transactions could be discarded to save disk space, but this may result in breaking within the blocks’ hashes and thus the immutability of a chain. This issue could be solved utilizing a Merkle tree, as explained in \cite{gupta2017blockchain}, in which the transactions are hashed with only the resulting root hash of such Merkle tree data structure stored in block headers respectively, the old blocks could then be compacted by stubbing off branches of the Merkle tree data structure involving transactions packed in a mined block.
  \item \emph{Privacy-Preserving}: The main benefit of using cryptocurrency is that identities involving the cryptocurrency are obfuscated thus anonymous. Even though a transaction is made public with its designated transaction hash is broadcast across the chain at first, privacy can still be preserved via breaking the flow of data, such as identified transaction senders and receivers, using secret public keys; for instance, entities on the network can see that a transaction of a specific amount is occurring between two specific entities but cannot map the transaction data to their individual identities. Each transaction is further protected by an additional firewall and a new key pair, but in some cases, such as multi-input transactions, some linking is unavoidable with the transaction inputs being owned by an identical owner. If a key of any owner of a digital asset is revealed, its subsequent links could be traced according to the transaction history and so the privacy could no longer be guaranteed. There are many privacy-preserving techniques technologies in place to enhance the ability of privacy preserving in different blockchain frameworks with different settings, such as \emph{Onion Routing} in \cite{syverson2001towards} and \emph{Garlic Routing} in \cite{dakhnovich2018applying} , where the former provides privacy features such as resisting traffic analysis and eavesdropping, and the latter could hide identities of the transaction packet sender and receiver utilizing layered encryption.
\end{itemize}

\subsection{Blocks}
A block consists of a batch of valid transactions with their corresponding transaction hash encoded into a Merkle tree, with the calculated root hash included in the corresponding block header. The uniqueness and immutability of blockchain technology is that the header of every block contains a cryptographic hash of the preceding block acting as pointers to form a chain, thus making it difficult to alter the data in any block as its validity could be traced back to the very first block (also known as the \emph{genesis block}), creating a link and ultimately creating a chain of such linked blocks. Each block has a block header that consists of a string, cryptocurrency version number, hash of the previous block, Merkle root of validated transactions, timestamp, data related to the mining difficulty, and the nonce which is the prime target of participating miners to initiate the block mining process of a network with consensus algorithm, such as \emph{PoW}.

Each block in the chain is worked upon by multiple participants or miner nodes on the network, and due to the presence of a specified algorithm for scoring different versions of the history (in addition to secure hash-based history), the one with the higher score is selected over others. The blocks not selected for inclusion in the canonical chains or any incomplete block, are known as orphan blocks. Though blockchain is an iterated process of forming a chain of blocks, under some specific conditions; for example, when two peer or miner nodes on the network find and broadcast a new block reference similar to that of the previous one in an environment with little to no permissions, a fork in the chain could be created and expected. The miner or peer nodes running on the same network may have different versions of history, such as the highest-scoring version known to them, recorded in their local chain data storage individually. Whenever a new version with the higher score is issued, such as the addition of a new block to the chain, these nodes will overwrite their own local storage individually, resend and broadcast these improvements to their peer miner nodes, as explained in \cite{1}.

\subsection{Block Time}
The amount of time required for the peer-to-peer network to solve and generate a new block on average, is known as block time. It is calculated based on the average amount of time required after recording data with a specific number of transactions validated with corresponding blocks mined, and such time window is assigned to miner nodes to avoid or monitor any potential security issues. Miner nodes working simultaneously on a single block could use high-performance systems that generate hashes at an astounding rate, and the block time could be as short as 5 seconds or even shorter depending on the chosen consensus algorithm of the blockchain network. This implies a fact that by the time a transaction takes place, its data becomes verifiable; hence, shorter block times implies faster transactions but could as well imply reduced security.

Other elements of block headers, such as the level of difficulty that differs for every mining process, could affect the block time of the mining process for which increasing difficulty level of puzzles (\emph{Difficulty Bomb}) in the mining algorithm would basically imply longer block time as there will be substantial lags between production of blocks on a specific blockchain. For instance, the difficulty bomb (also known as the "\emph{Ethereum's Ice Age}" of its original PoW implementation on Ethereum main net) was introduced for the transition of Ethereum Proof-of-Work to \emph{Proof-of-Stake} on its main net termed as \emph{Eth 2.0} as depicted in \cite{saleh2018blockchain}, by increasing the difficulty level on its PoW mining algorithm to be more complex leading to longer block time and less appealing to miner nodes participating in PoW mining algorithm of Ethereum main net. The average block time for Bitcoin transactions is roughly 10 minutes.

\section{The Block Mining Process}
The process of validating blockchain transactions with the transactions packed in a block, as described in \cite{eyal2014majority}, is generally termed as mining, and this process is used to manage cryptocurrencies and other digital assets, as well as rewarding miners based on their computational contribution, on a decentralized and peer-to-peer blockchain network. This mining process could further be elaborated with the example of Bitcoin, which is the first and primary cryptocurrency adopted in the blockchain field, for the purpose of trustless peer-to-peer electronic payments. Cryptocurrencies have not yet been formally regulated by general governments and have been designed with absolutely no need of involving any intermediaries, so how does it promote trust amongst users to adopt it for different financial purposes, and how could these be considered valid and secure with finality and non-repudiation guaranteed?

Similar to any fiat currency circulating in any financial system worldwide - which by itself is of no value but represents the trust that the users of those currencies put into their respective government and economy to uphold any monetary transaction - digital and cryptocurrencies are managed by anonymous but multiple (in millions) crypto miners and digital communities across the globe, and the process itself is protected by cryptography techniques in place, which is also iterative, distributed and decentralized. Thus, blockchain could be described as the process of building trust between trustless parties via trusting the shared code and signature system applied to the blockchain network via the corresponding blockchain nodes.

Bitcoin transactions are recorded in the form of hexadecimal value blocks, and the aim of the crypto miners is to be rewarded via solving these blocks of peer-to-peer transactions utilizing mathematical calculations to verify their authenticity. For mining one block of size 1 MB, Bitcoin rewards the miner (usually the first miner node to solve the hash of a mining round) who spent computation resources verifying transactions during the block mining process so as to keep the integrity of Bitcoin network, with a specific amount of cryptocurrency from a portion of transaction fee settled by the transaction senders involved. The block mining process, performed by the miner nodes, will, in turn, facilitate the Bitcoin network to mint more cryptocurrency over time for circulation. At its inception, solving each Bitcoin block came with a reward of 50 BTC, which have been halved at every specific milestone block predefined, hence, known as \emph{Bitcoin Halving} as elaborated in \cite{devries2016analysis}; in May 2020, the reward for solving one block is \emph{6.25 BTC}. This process of such extreme competition, on winning the block, involves many miner nodes with innumerous computational resources required. From the perspective of cybersecurity, such a lucrative offer of rewarding cryptocurrency to the block winner (the first miner node solving a block) has attracted many malicious actors, and has introduced a concept of mining pools which could in turn enhance the degree of centralization on the entire blockchain network.

To mine the block quickly and to be more likely to include blocks on the canonical chain of the network, miners either need to join mining pools to contribute their computational power so as to get reward proportionally only if the mining pool successfully wins the block later, or possess with  high-performance units with their designated nodes running on the network which would normally require a memory-hard requirement for which every Bitcoin miner forced using chips that are highly optimized for computing hash values, such as the specified \emph{SHA25} hashing algorithm of Bitcoin mining. While legitimate miners use their own high-end systems and tools, hackers tend to go around these investments and hack devices to serve as bots for them. These computational resources could then be combined together (in thousands) to form a mining pool and to perform the mining tasks of generating and throwing out hashes by millions at the block. On the contrary, Ethereum \cite{wood2014ethereum} introduced \emph{GHOST} (Greedy Heaviest Observed Subtree) protocol allowing miners to get extra rewards if their mined stale blocks are included with any successful block included on the canonical chain. This would mean improved security and more profit for independent miners contributing to the security of the network as it becomes more decentralized, compared with the mining pools of the Bitcoin network centralizing hash power and resources of the network. Nonetheless, Ethereum dismisses a memory-hard requirement for the participating miners to encourage miners using affordable and general GPU instead of relying on \emph{ASIC} (Application-Specific Integrated Circuit) producers, as explained in \cite{taylor2017evolution}, and thus preventing ASIC producers and miners from forging a 51\% share or attack to Ethereum network.

\section{Blockchain Coordination}
The blockchain network, being run and joined by distributed nodes maintained by decentralized institutions and entities, implements transparency of operations and functionalities. Blockchain network could be generally categorized as \emph{permissioned} or \emph{permissionless} (public blockchain), where the latter is not governed by any institution which could raise concerns over the existence of incomplete contracts and difficulty of policy incorporation and implementation, such as retracting a compromised chain or currency exchange rate of on-chain digital assets, etc. Unlike Bitcoin and Ethereum, which rely on developer communities for core policy decisions, recent projects, such as \emph{EOS} as elaborated in \cite{xu2018eos}, have a formal on-chain governance process, which highlights the need of continued innovation in blockchain governance, with research effort performed on economic design for decentralized governance systems to improve the outcome. Incorporating such designs of decentralized governance into the blockchain process could introduce new features such as voting to decide if a hard fork should be planned, which could make the miners and participants of a blockchain network feel sufficiently effective due to the fact that such scenarios could be extremely costly in some contexts. Many blockchain experts currently argue about the benefit of hard forks and their corresponding implementations according to different circumstances. In such instances, the welfare economic tools and designs can be utilized to estimate the benefit of hard forks under those particular conditions, along with depicting its maximum efficiency and ideal environment as discussed in \cite{3}.

Another feature or benefit of economic welfare design is to manage network effects, where the number of nodes on a chain, or the size of the blockchain network could define the value of that chain to miners and participants of the blockchain network. As such, when defining policies in a governance setting, opinions are subjected to the preferences of miners and participants on that chain. The common voting procedures do not properly work or may face unexpected and detrimental outcomes in the presence of both hard forks and network effects in a blockchain setting. Studies have proposed solutions to solve or overturn decisions of the majority through quadratic voting in the instance of a situation where the passionate minority group is not in agreement with the majority’s preferences, and hence causing a hard fork. Although this would not cause a major issue as the hard-forked chain has owned its preferred policy, the network effects could cost a large sum borne amongst the community. Thus, the need of a decentralized governance system is important in coordinating the community on a single policy.

Some studies have found that it is always a \emph{Nash equilibrium}, as explained in \cite{holt2004nash}, for everyone to choose to stay on the main chain or the upgraded chain, depending on the composition of the user base and the relative benefits. There are always some parameters that could maximize the social welfare factor of hard forks, but it may or may not increase the total surplus of the blockchain community. Neither the majority rule nor the quadratic voting, which could grant the benefit to a minority group, could eliminate sub-optimal hard forks, and the only solution is to avoid altering the proposal-making process altogether. Though decentralized governance systems work well with policy formulation and implementation, there is a need for the blockchain developers and corresponding communities to widen their definition of decentralized governance systems during the stages of framework design and development to include voting, proposal-making, and communications operations on the general blockchain network and any corresponding blockchain client implementations for miners and participants running blockchain nodes on the network.

\subsection{Smart Contract At The Core}
Smart contracts are self-executing scripts with predefined methods deployed and stored on blockchain available to participating nodes running on the network to access via a designated blockchain client. Smart contract methods are invoked to execute accordingly on every node of the blockchain network when performing a transaction. Every node of the blockchain network must agree on the inputs, outputs and states transited by the smart contract. Common contractual conditions, such as payment terms and conditions could be satisfied, thus minimizing the need for trusted intermediaries but the decentralized "\emph{code}" instead.

For instance, the agreement terms binding the buyer and seller, in any standard transaction, could be translated and written into lines of code — thus making the code that contains the agreement of the contract available across the decentralized and distributed, public or permissioned, blockchain networks. The complete agreement in the form of the transaction cannot be modified and is traceable, and therefore such features of smart contracts could help and permit trusted transactions and agreements to occur between anonymous parties without the involvement of government authorities, legal systems, external enforcement or with any form of intermediaries.

\subsection{Smart Contract Over Different Blockchain Implementations}
Blockchain has evolved from merely having a use case of being a peer-to-peer network to process cryptocurrency transactions, where participating nodes enact consensus protocols, to function as state machines that consistently track transitions to data and states jointly manipulated by a number of clients. This tamperproofing sequence of blocks and the anonymity around it has attracted entities and users willing to share consent between mutually-untrusted parties. This blockchain feature has enabled a generally-purposed and distributed computing platform. Blockchain technology helps implement smart contracts in multiple ways, such as implementing stricter admission rules for new agents, thus allowing for finer access control and the concept of \emph{peer-permissioning}. Some of the prime blockchain implementations and their interactions with smart contracts are described below.

\begin{itemize}
  \item \emph{Ethereum}: \cite{wood2014ethereum} is a permissionless blockchain implementation that allows both node accounts and smart contract accounts to hold balance in Ether - the native cryptocurrency of Ethereum implementations. The smart contract encapsulates custom and stateful behaviors, for which miner nodes could run the hexadecimal bytecode data of the transaction with its \emph{Ethereum Virtual Machine} (EVM), which is the execution environment for running transaction code to reach a consensus. Smart contracts are compiled with EVM to generate operation codes (e.g. \emph{PUSH1 0x60 PUSH1 0x40 MSTORE}) and run based on the operation codes each time a specific method in the smart contract related to a specific transaction is invoked following the rules set within the deployed smart contract. The smart contract could then have states transitioned on each miner’s local persistent storage on state data, only if data of the transaction is executed by the EVM successfully. There are also a variety of transaction types in Ethereum, namely the deployment transaction to deploy the new smart contract to the blockchain network, cryptocurrency transfer transaction to exchange cryptocurrency and update balances of different node accounts, and invocation transaction to send messages to smart contracts. Ethereum does not encourage complex computation steps and long computation time in smart contracts through the concept of fuel included in the transaction called "\emph{Gas}". There is also a permissioned implementation of Ethereum with a consensus algorithm of \emph{PoA} (Proof-of-Authority) with peer-permissioning and on-chain governance enabled, which are covered in \cite{2, 4}.
  \item \emph{Hyperledger Fabric}: \cite{androulaki2018hyperledger} is a permissioned blockchain implementation in which nodes consisting of "\emph{Ordering Service}" are in charge of a consensus protocol. The roles of the miners are assigned by the "\emph{Membership Service}", along with their membership verification and nodes that have access to the ledger are called "\emph{peers}", and each peer belongs to some organizations. Smart contracts could be dynamically installed and executed in any given channel.
  \item \emph{Corda}: \cite{hearn2016corda} is a permissioned blockchain framework, which is made of different nodes where each is identified by authority only capable of issuing certificates, termed as "system doorman service". Corda is designed with privacy as its primary concern, and thus it is devoid of shared state of data, and each node gets only a portion of the system state, called Vault. The smart contract is called Contracts and is deployed by the system administrator, who could check the validity of transactions.
  \item \emph{Tenderfone}: As explained in \cite{ciatto2020agents}, Tenderfone is a permission blockchain implementation with proactive smart contracts that are expressed via a language based on \emph{First Order Logic} (FOL). The smart contract in Tenderfone consists of logic facts according to the data in it containing logic terms that can virtually represent arbitrarily complex data structures. The smart contract computes through logical reasoning and provides common goals through deduction, which is either furnished during the smart contract creation or could be provided by both off-chain users or other smart contracts through messages.
\end{itemize}

\section{Blockchain Forks}
A fork is a terminology used to signify the cutting off or the diversion from the mainstream or existing technology, framework, policy, environment, etc. Blocks of the Bitcoin network are generated in the process of mining with validated transactions included in a block and solving a Proof-of-Work puzzle for that block. A blockchain fork occurs when two miners find and publish a new block referencing independently, where the reference is similar to that of the previous block. Though the inconsistency is temporary and is resolved by the subsequent blocks, forks shift the existing rule towards a new set of predetermined rules and are a part of the blockchain’s normal operation. Reasons such as the existence of delay, deviating mining strategies, the difference in policy opinions which leads to blockchain forks, where the latter, also known as negative gamma, affects the security of blockchain implementations along with speculations of network topology in a permissionless environment. The occurred forks, as covered in \cite{karame2016security} generally help to understand the factors that originally led to its formation. Generally, the main cause for a fork is when two independent miners solve the Proof-of-Work puzzle at the same time, but it can also arise from selfish mining for which the miners solve and withhold new blocks instead of immediately broadcasting them in order to gain an advantage in finding the next block. The new blocks are propagated by flooding them on the peer-to-peer network, or by transmission via additional, possibly private networks as explained in \cite{5}.

\subsection{Hard Fork}
If an upgrade in the blockchain process is non-backward compatible with the existing blockchain implementation, it is termed as a hard fork. It can result from a rule change such that the client software and virtual machines validating transactions and blocks according to the old rule will consider it invalid and redundant, hence requiring upgrades of all the nodes involved. One of the prime examples of this is the hard fork of Bitcoin and Bitcoin Cash. Imagine in a blockchain process, after the nth block, some rules in the node were changed, or an upgrade in the blockchain protocol was proposed. The miner nodes (peers on the peer-to-peer network) are required to vote on the adoption and incorporation of these changes. In case the incumbent process is elected, no changes would occur, but if the upgrade is elected and made, it could trigger a "\emph{hard fork}" in the chain. For instance, any block mined using old protocols would become invalid and could not be chained to the block mined with the new protocol as depicted in \cite{6}.

\subsection{Soft Fork}
Unlike the hard fork, a soft fork is a backward-compatible upgrade, where the upgraded nodes are able to communicate with those non-upgraded nodes. This is possible only if the new and upgraded rule does not clash with the old one. Examples of this can be seen in both Bitcoin and Ethereum blockchains, where the published and incorporated upgrades are backward-compatible. This process also leads to deviation in the chain sometimes when any particular block mined using old rules is not processed by the upgraded nodes, but the non-upgraded nodes process it instead - leading to unsynchronization without the users noticing it.

\subsection{Impact of Blockchain Fork}
The extent of the effect of any hard or soft fork on the blockchain is dependent on the environment and other variables involved. In a hard fork, the nodes will be impacted severely to the extent of block invalidation. This implies that if a block or node is not upgraded, the rules do not match the upgraded counterparts, and the block is rendered invalid. The changes in the hard fork cannot be undone or deleted after the fork is completed, but the opposite is possible in the soft fork; for example, if any old node decides to make and verify a block, it ends up being considered valid by all other nodes on the network, irrespective of their upgrade status as explained in \cite{7}. The entities and institutes of the blockchain process that get affected by forks are:

\begin{itemize}
  \item \emph{Miners}: The crypto miners or peers are the main force involved in the decision for the adoption and incorporation of any upgrade process or fork. They need to decide as to whether the incorporation of the \emph{Blockchain Improvement Proposal} (BIP) will maximize the functionality and their revenue along with other benefits.
  \item \emph{Hardware Manufacturers}: The change in algorithm and software sometimes involves modifications in the hardware completely due to the fact that the hardware could be incapable of supporting the proposed change. This might cause loss to the hardware clients and manufacturers. Even if new hardware is proposed for development, the manufacturers and their development team will be required to plan, design, and test the hardware for functionality and security, involving a large investment of resources and time with little or no direct gains.
  \item \emph{Developers}: Developers could also be considered as a resource. If a fork arises due to development issues or software bugs, the human resource is bound to get divided while creating vacancies that may not be easily filled, hence affecting the service and operations.
  \item \emph{Market Exchange}: The fork does not divide the number of digital assets an address holds, but it could divide the value of the digital assets or cryptocurrencies. For example, if a user/address holds 2 BTC before the fork (Bitcoin to Bitcoin Cash),  the value after the fork remains 2 BCH (Bitcoin Cash) but the values between the same amount of Bitcoin and Bitcoin Cash are very different in today's crypto market. It is up to the exchanges to build capabilities such that the user gets tokens allotted on the new or old fork. The contentious hard fork generally airdrops the public and private key pair with the same number of native tokens to the new chain as in the old ones.
  \item \emph{Investors}: The market is highly volatile when dealing with investment in forked commodities. This turmoil induced by a contentious hard fork gives rise to a wild swing in prices. There is also uncertainty in new forks or sometimes in old forks depending on the public and market opinion. This is one of the factors blockchain communities need to consider while deciding on the BIP which would later lead to blockchain forks.
  \item \emph{Merchants}: Merchants need to set up the new infrastructure in compliance with the new fork in order to accept the new cryptocurrency initiated from the new fork.
  \item \emph{Wallet Developers}: Similar to merchants, digital wallet developers need to plan and restructure their applications to incorporate and maintain crypto assets of the new blockchain.
\end{itemize}

\subsection{History of Blockchain Fork - Bitcoin}
\begin{figure}[h]
    \centering
    \captionsetup{justification=centering}
    \includegraphics[width=0.5\textwidth]{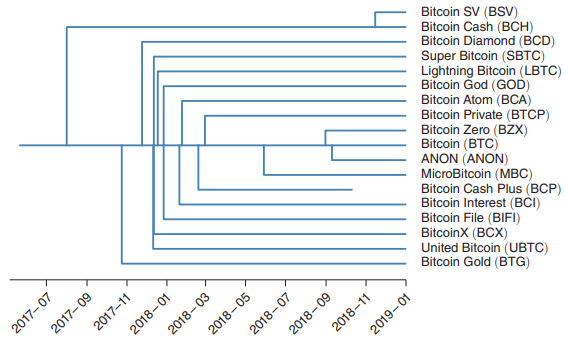}
    \caption{\textit{Fork Events of Bitcoin Network, Source: Biais, et al.}}
    \label{fig:forkevent}
\end{figure}

With the advent of Bitcoin Network released back in 2009, there has been a wide range of fork events took place over the year on the Bitcoin Network as summarized in \textit{Fig.~\ref{fig:forkevent}} .In June 2014, Mike Hearn published a Bitcoin Improvement Proposal (BIP 64) that called for the addition of a P2P protocol extension to perform \emph{UTXO} look-ups. An upgraded version (0.10) was issued in December 2014 (forked client Bitcoin XT) that attracted significant attention within the Bitcoin community. In 2015, Gavin Andresen published another improvement proposal (BIP 101), calling for an increase in the maximum block size to 8 MB and a subsequent increase by two, every year, once 75\% of a stretch of 1,000 mined blocks was achieved. Furthermore, the maximum transaction rate was 24 transactions per second. In August 2015, BIP 101 was merged into the Bitcoin XT code base, and a 2 MB size bump of Bitcoin Classic was applied. In August 2017, Bitcoin XT published Release G, a client of Bitcoin Cash, as discussed in \cite{8}.

The Bitcoin network protocol initially set the size of its block to 1 MB, but with the increase in the Bitcoin protocol on its blockchain, it was compelled to upgrade its protocol to increase throughput. In May 2017, developer and miner communities came up with an agreement namely the "\emph{New York Agreement}" (NYA) in order to facilitate the upgrade under the \emph{SegWit2x} project, where the agreement planned a soft fork (SegWit upgrade) of increasing the block size to 2 MB by November in that year. However, an alternative hard fork, Bitcoin Cash, occurred in August 2017 that was supported by Bitmain, a major manufacturer of ASIC mining software and tools, and other mining pool operators. Bitmain owned a patent on \emph{AsicBoost} (a mining-enhancing technology) that had limited use with the SegWit planned to be used by the New York Agreement as elaborated in \cite{9}.

Another important hard fork was Bitcoin Gold in 2017. It came about on the argument that the network and the original decentralized structure were threatened by the dominance of \emph{ASIC} manufacturers, who could compel the miners to follow their terms. Thus, Bitcoin Gold would prevent the use of ASICs, and help restore the Bitcoin blockchain to its original structure. However, this fork received some criticism as the developers of Bitcoin Gold pre-mined 100,000 coins before opening the blockchain to the wider public and also included a hidden fee of 0.5 percent of all block rewards in its code proposed to set up Bitcoin Gold mining pools. The hidden fee was later removed by public miners with designated nodes running on the blockchain network.

\section{Conclusion}
Blockchain technology is widely known for its use case of digital currency with secure and transparent data integrity related to peer-to-peer digital payment transactions. The decentralized and distributed characteristics of  Blockchain technology have made tampering of data and states difficult while building a community around itself through incentives that generate and popularize cryptocurrency and any blockchain-enabled digital assets. The functionality of blockchain is no longer limited to the primary use case of cryptocurrency, but can also be implemented across different service sectors where security and anonymity of transactions and communications in the digital space are in high demand. The smart contract is one such example of implementing the concept of blockchain technology in the legal and contract binding process. The forks that arise in blockchain technology and in any current blockchain implementation could be influenced by a variety of factors, and require the decentralized consensus to come into existence, which could end up creating a completely new process, or new blockchain implementation with a native cryptocurrency.

\bibliographystyle{ieeetr}

\begin{IEEEbiography}[{\includegraphics[width=1.1in,height=1.25in,clip,keepaspectratio]{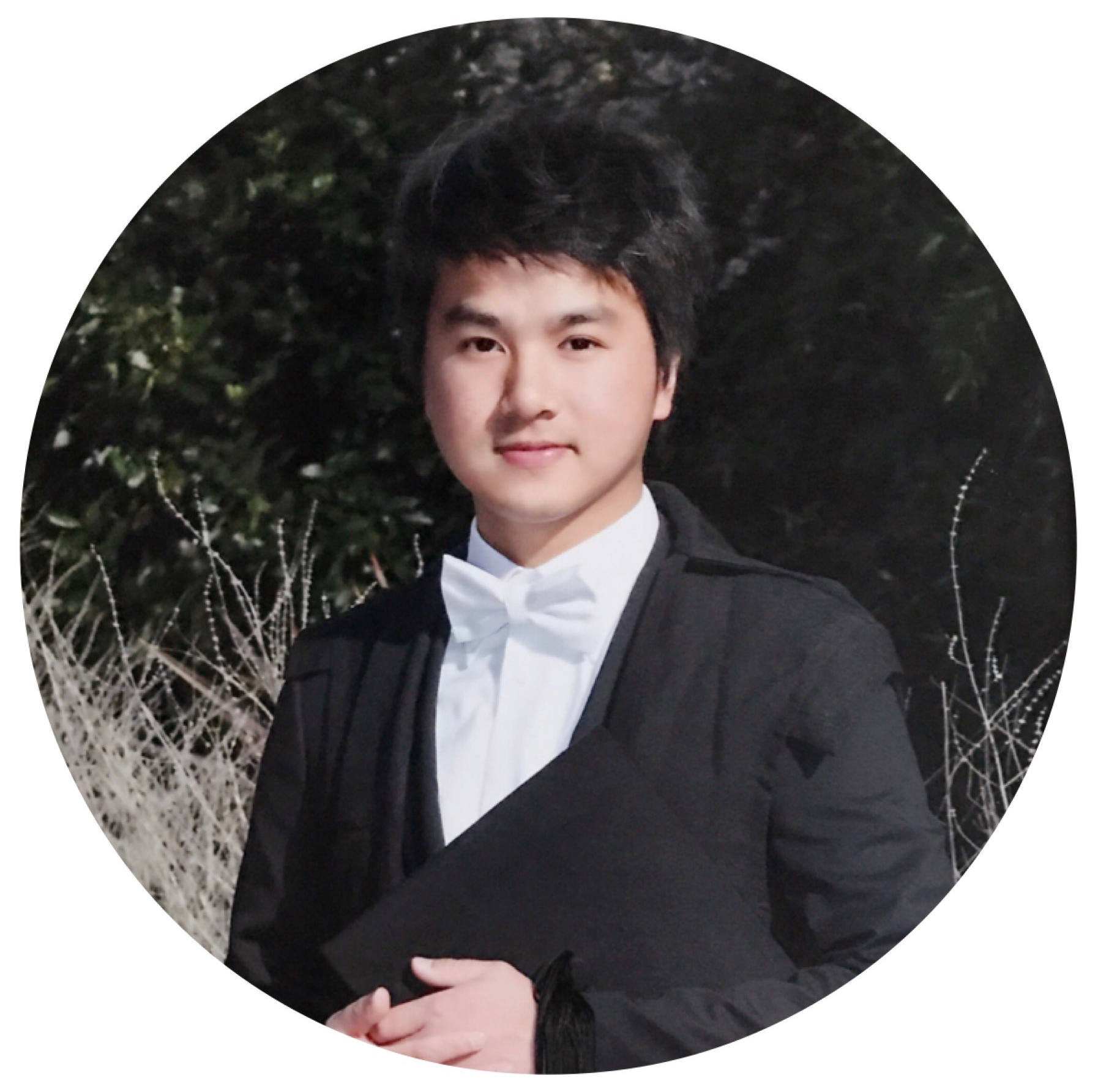}}]{Mr. Neo C.K. Yiu IEEE}
is a computer scientist and software architect specialized in developing decentralized and distributed software solutions for industries. Neo is currently the Lead Software Architect of Blockchain and Cryptography Development at De Beers Group on their end-to-end traceability projects across different value chains with the Tracr™ initiative. Formerly acting as the Director of Technology Development at Oxford Blockchain Society, Neo is currently a board member of the global blockchain advisory board at EC-Council. Neo received his MSc in Computer Science from University of Oxford and BEng in Logistics Engineering and Global Supply Chain Management from The University of Hong Kong.
\end{IEEEbiography}
\vfill
\end{document}